\documentclass[prd,preprint,eqsecnum,nofootinbib,amsmath,amssymb,
               tightenlines,dvips]{revtex4}
\usepackage{graphicx}
\usepackage{bm}
\usepackage{bbm}
\input epsf

\begin{document}

\def\centerbox#1#2{\centerline{\epsfxsize=#1\textwidth\epsfbox{#2}}}
\def\Eq#1{Eq.~(\ref{#1})}
\def\AL{A_{_{\rm L}}}
\def\FL{F_{_{\rm L}}}
\def\CL{C_{_{\rm L}}}
\def\i {{\mathbbm i}}
\def\be{\begin{equation}}
\def\ee{\end{equation}}
\def\baray{\begin{eqnarray}}
\def\earay{\end{eqnarray}}
\def\lsim{\roughly<}
\def\gsim{\roughly>}
\def\ve{\overrightarrow}
\def\hf{\frac{1}{2}}
\def\Str{\text{Str}\;}
\def\Tr{\text{Tr}\;}
\def\tic{$\surd$}
\def\X{$\times$}
\newcommand{\rphi}[1]{\Re\left[\phi\left(#1\right)\right]}
\newcommand{\iphi}[1]{\Im\left[\phi\left(#1\right)\right]}

\title{Defect Formation with Bulk Fields}

\author{Guy D.\ Moore and Horace Stoica}
\affiliation
    {%
    Department of Physics,
    McGill University, 3600 University St.,
    Montr\'eal QC H3A 2T8, Canada
    }%

\date{\today}
\begin{abstract}{
It has recently been realized that brane-antibrane annihilation (a
possible explanation for ending inflation) may result in defect formation, due
to the dynamics of the tachyon field.  Studies of this possibility have
generally ignored the interaction of the brane fields with fields in the
bulk; recently it has been argued \cite{Dvali:2003zj} that interactions
with bulk fields suppress or even eliminate defect formation.

To investigate the impact of bulk fields on brane defect formation,
we construct a toy model that captures the essential features of the
tachyon condensation with bulk fields.  We study the structure of
defects in this toy model, and simulate their formation and evolution on
the lattice.  We find that, while the energetics and interactions of
defects are influenced by the size of the extra dimension and the
bulk-brane coupling, the bulk-brane coupling does not prevent the
formation of a defect network. }
\end{abstract}

\maketitle

\section{Introduction}

The early universe is the most likely source for experimental input into
physics at very high energies.  Immediately after the Big Bang, energies
as high as the Grand Unified (GUT) scale may have been explored.  It is
hoped that signatures of the processes that took place at that time
will be visible in the universe today, for instance, in primordial
perturbations or other relics of the very early universe.

Recently models for the early universe have become less ad-hoc
and more inspired from String Theory \cite{BraneInflation}.
The brane world model is particularly attractive,
since it contains the ingredients for a successful inflationary universe.
The separation of a brane and an antibrane can act as an inflaton field,
and the interaction
of the pair allows one to calculate the corresponding inflaton potential.
The collision and annihilation of the brane-antibrane pair provides a
natural mechanism for ending inflation and reheating the universe.

One important prediction of a large class of brane inflation models is
the formation of topological defects, typically networks of cosmic
strings---it has been argued \cite{Tye:StringProduction} that
brane-anti-brane annihilation
in the very early universe leads to the formation of cosmic strings
only [see also \cite{Barnaby:2004dz}].
The formation of topological
defects in brane-anti-brane annihilation has been well studied in String
Theory and Field Theory
\cite{CosmicStrings,StaticKinks,DescentRelations}, and represents a
particularly testable potential signature for these models, since the
defect network can have quite different properties from that of
conventional cosmic string networks \cite{Tye:StringProduction}.

The discovery of explicit mechanisms for the stabilization of the
compact extra dimensions makes this class of models even more appealing
for phenomenology \cite{ModuliStabilization}.
These types of models have been mainly used to describe the
inflationary universe, and the predictions these models make for the
cosmological parameters can already be tested against the experimental data. At the same time,
the mechanism that stabilizes the compactification poses additional challenges since it
can make the topological defects unstable. \cite{CosmicStrings}

The formation of cosmic strings in brane-anti-brane annihilation has
also been studied in the
4D low-energy effective theory, by analyzing the dynamics of the fields
living in the world-volume of the brane \cite{StaticKinks}.
One usually considers only the complex tachyon field or can include the
$U\left(1\right)$ gauge field that couples to the tachyon. Bulk fields
are usually neglected.  However,
it has been argued in \cite{Dvali:2003zj} that the inclusion of the bulk
fields in the process of defect formation substantially changes the
dynamics of defect formation, leading to a strong suppression
of the defect formation rate.  This is because the topological defects
are themselves branes, so they couple to fields described by closed
string modes, namely the
dilaton and the Ramond-Ramond fields which propagate in the entire
bulk space-time
\cite{Polchinski:1998rq,Polchinski:1998rr,Johnson:2000ch,Douglas:1995bn}.
The energy of the gradients of the RR field
along the compactified extra dimensions is large, so the formation of
defects is strongly suppressed. 
More precisely, it was argued that deSitter quantum fluctuations
cannot account for the formation of lower dimensional branes since the
typical energy of the quantum fluctuations is of order the Hubble
constant $H$ while the gradients of the fields spreading in the extra
dimensions of size $R$ are of the order $1/R$. During brane inflation
$H \ll 1/R$, so the formation of lower dimensional branes is
suppressed.

In this paper we try to analyze the mechanism of defect formation when
bulk fields are involved.
We first construct a toy model which captures the essential features
of the full 10-dimensional model that are responsible for the formation of the
topological defects.
Using this toy model we study the classical evolution of the brane and bulk
fields, so the deSitter quantum fluctuations are responsible only for
seeding perturbations which grow during the subsequent evolution. The final
configuration of the bulk field gradients is a result of this
classical evolution.  The energy of
the system comes from tension of the brane-anti-brane pair, which 
can be modeled as the potential energy of the tachyon field sitting at
the top of the potential. In
our toy model this corresponds to the potential energy of a complex
scalar living in the worldvolume of the brane-anti-brane pair.

We will present the SUGRA Lagrangian for the fields which live both
inside the worldvolume of the brane-anti-brane system and the bulk, and
we will construct a toy model with the same physical features.
We will evolve the equations of motion of this toy model on a lattice
and observe the formation of the defects. We will analyze the effect the
coupling to the bulk fields has on the formation process as well as the
characteristics of the defects once they are formed.

In section 2 we present the relevant supergravity Lagrangian and the toy
model we use.  In section 3 we analyze the equations of motion of the
toy model and discuss the energetics for a defect-anti-defect pair. In
section 4 we describe the results we obtained from lattice
simulations. Finally the conclusions are presented in section 5.
Details of the lattice implementation of our toy model are left to a
technical appendix.

\section{Tachyon condensation with bulk fields}

The usual approach to defect formation during tachyon condensation is to take into account only
the fields that live in the worldvolume of the decaying non-BPS brane or brane-anti-brane pair.
This approach seems to be motivated by the fact that the resulting lower-dimensional branes
formed during the decay are localized inside the worldvolume of the
parent brane. However, the final state defects are themselves D-branes,
and therefore couple to bulk RR-fields.  One
should include the effects of these fields in the defect formation process.

When studying the formation of topological defects by the brane
worldvolume fields only, one
usually considers a single non-BPS brane with the action:
\be
\label{single_non_BPS}
S=-T_{p}\int d^{p+1}x \; e^{-\phi}\:V(T)
 \sqrt{\det\left(P\left[G_{ab}+B_{ab}\right]+ 
2\pi\alpha^{\prime}\left[F_{ab}+\partial_{a}T\partial_{b}T\right]\right)}\,,
\ee
where $P\left[G_{ab}+B_{ab}\right]$ represents the pull-back of the bulk metric and
NS-NS two-form field on the brane and $F_{ab}$ is the field strength
of the Abelian world-volume gauge field \cite{DBIAction,DBIderive,DBIproperties}.
This action was used extensively to study the evolution of
the tachyon field in various special settings. The most common one is the spatially
uniform field in a Friedmann-Robertson-Walker universe. In this case one usually sets the
NS-NS two-form field and the brane gauge field to vanish everywhere, and study the time evolution
of the tachyon field and scale factor of the universe. The uniform field does not 
lead to the formation of defects, it behaves like
a pressureless fluid known as ``tachyon matter'' \cite{TachyonMatter}.

When studying the formation of topological defects during tachyon condensation, one usually
chooses a flat metric, sets the NS-NS field and the brane gauge field to vanish everywhere, but
chooses a tachyon field that is both time and space dependent. The equation of motion for
the tachyon field,
\be
\frac{\partial_{\alpha}\left(\sqrt{-g}g^{\alpha\beta}\partial_{\beta}T\right)}
{\sqrt{1+\partial_{\alpha}T\partial^{\alpha}T}}
-\frac{\sqrt{-g}g^{\alpha\beta}\partial_{\beta}T\partial_{\alpha}
\left(\partial_{\mu}T\partial^{\mu}T\right)}
{2\left(1+\partial_{\alpha}T\partial^{\alpha}T\right)^{3/2}} -
\frac{V^{\prime}\left(T\right)}{V\left(T\right)}
\frac{\sqrt{-g}}{\sqrt{1+\partial_{\alpha}T\partial^{\alpha}T}} = 0 \, ,
\ee
is non-linear, and one usually solves the equation around a point where
$T = 0$. There
the space and time dependence can be approximated by a linear space profile
with a time-dependent slope,
$T\left(t, x\right)\simeq u\left(t\right)x$, and the resulting equation for
$u\left(t\right)$ can be solved. The solution becomes singular in finite
time, the occurrence of the singularity marking the formation of the
topological defect. This result
confirms the String Theory calculation in which a linear tachyon profile
$T\left(x\right) = ux$ reproduces the correct tension of a codimension
one brane in the limit $u\rightarrow\infty$.

The brane gauge field is usually included in the form of a uniform background electric field
\cite{Mukhopadhyay:2002en} or a constant gauge potential (Wilson line) \cite{Hashimoto:2002xt}. 
The case where both the brane gauge field and a worldvolume 
scalar field other than the tachyon are included was studied in \cite{Mehen:2002xr}.

This action is highly non-linear and in order to perform a lattice regularization we prefer
an expression in which the square-root is expanded to quadratic order in the field strengths.
Also for a single non-BPS brane the tachyon is a real scalar which cannot be minimally coupled to
the world-volume gauge field.

It is therefore more convenient to consider the action for a brane-anti-brane
pair. In the case of a $D9-\overline{D9}$ pair the expanded action involving the complex
tachyon coupled to the gauge fields living inside each brane is given in Ref. \cite{Kraus:2000nj}:
\baray
S&=&2T_{D9}\int d^{10}x \, e^{-\phi} 
 e^{-2\pi\alpha^{\prime}T\overline{T}}\left[
1+8\pi\alpha^{\prime}\ln\left(2\right)D^{\mu}\overline{T}D_{\mu}T+
\phantom{\frac{\left(2\pi\alpha^\prime\right)^2}{8}} \right.
\nonumber \\
&& \left.\qquad \qquad \frac{\left(2\pi\alpha^{\prime}\right)^2}{8}
\left(F^{+}_{\mu\nu}\right)^2+
\frac{\left(2\pi\alpha^{\prime}\right)^2}{8}
\left(F^{-}_{\mu\nu}\right)^2+
\frac{\beta{\alpha^{\prime}}^2}{8}
\left(F^{+}_{\mu\nu}-F^{-}_{\mu\nu}\right)^2 \right]\,.
\earay
The two gauge fields live in the worldvolume of each brane and the
tachyon field couples only with one linear combination:
\be
D_{\mu}T = \partial_{\mu}T-\left(A_{\mu}^{+}-A_{\mu}^{-}\right)T \, .
\ee

The brane fields couple with the Ramond-Ramond (RR) bulk fields through the Chern-Simons
coupling also given in Ref. \cite{Kraus:2000nj}:
\be
S^{D\overline{D}}_{RR}=T_{D9}\int C\wedge \Str e^{2\pi i
  \alpha^{\prime}{\mathcal F}} \,.
\ee
One can expand the exponential above, and the leading order coupling
between the brane fields and the bulk RR field involves the same linear
combination of brane gauge fields that couples to the tachyon:
\be
S^{D\overline{D}}_{RR}=2\pi \alpha^{\prime}T_{D9}
\int C_{p-1}\wedge \left(F^{+}-F^{-}\right)
\ee
We see that the orthogonal linear combination, $A_{\mu}^{+}+A_{\mu}^{-}$ does not
couple to any other fields, so we will drop it from the action.
Including the kinetic terms for the RR field, the dilaton and the metric,
the 10-dimensional action describing the decay of the brane-anti-brane pair is:
\baray
&& S=\frac{1}{2\kappa_{10}^2}\int \sqrt{-G} \: d^{10}x \left[
e^{-2\phi}\left(R+2\left(\nabla\phi\right)^2\right)-\frac{1}{2\left(p\right)!}F_{p}^2
\right]
\nonumber \\
&& -2T_{D9}\int d^{10}x \,e^{-\phi}\, 
e^{-2\pi\alpha^{\prime}T\overline{T}}\biggl[
1+8\pi\alpha^{\prime}\ln\left(2\right)D^{\mu}\overline{T}D_{\mu}T+\biggr.
\biggl.\frac{\left(2\pi^2+\beta\right){\alpha^{\prime}}^{2}}{8}
\left(F^{+}_{\mu\nu}-F^{-}_{\mu\nu}\right)^2\biggr]\,, \quad \quad
\earay
where $F_{p}$ is the corresponding field strength for the potential $C_{p-1}$, $F_{p}=dC_{p-1}$.
The $C_{p-1}$ will be the only field we consider here, we will not include the
dilaton and the metric in the toy model we consider.

\section{A toy model}

We want to create the simplest toy model which captures the important features of the formation of
defects with bulk fields included, and at the same time is amenable to a lattice regularization.
This will allow us to follow the evolution to see whether defects form,
and how the interactions with the bulk fields affect that formation.
We want to include the minimal field content that will allow us to study
the formation of the defects, so we will not include the metric and
dilaton fields present in the full 10-dimensional model.

Since we want to study the formation of codimension 2 defects the brane
must have at least 2 spatial dimensions, so we will choose a 2+1
dimensional brane. The defects that form are 0+1-dimensional 
and the corresponding bulk field that couples to their worldvolume must
carry a single index.  Therefore we will have a vector field living in the bulk.
This field corresponds to the $C_{p-1}$ RR field in the 10-dimensional model.
We choose the bulk to have the minimal space dimensionality, 1 space dimension more than
the brane. Inside the brane we put the same field content as in the full 10-dimensional model,
an Abelian gauge field corresponding to the linear combination $F^{+}_{\mu\nu}-F^{-}_{\mu\nu}$
and a complex scalar field corresponding to the complex tachyon $T$.

Therefore the model consists on a complex scalar $\phi$ charged under a gauge field $A_{\mu}$ living
in a 2+1 dimensional ``brane'' (see Table \ref{toy_model}). 
As in the full String Theory case, the fields have a Chern-Simons
coupling to the 3+1 dimensional bulk gauge field $C_{\mu}$. For our purposes it is sufficient to
keep only the coupling of the brane gauge field to the bulk field and ignore the coupling
of the scalar, which is higher order in the string coupling,
as we discuss in Appendix \ref{appA}.

\begin{table}[tbh]
\begin{center}
\begin{tabular}{|c|c|c|c|c|}
\hline
 & \multicolumn{2}{c|}{10D model} & \multicolumn{2}{c|}{toy model}  \\
\hline
 & dimension & field content & dimension & field content \\
\hline\hline
Brane  & p+1 & $A_{\mu}^{+}-A_{\mu}^{-}$, $T$ & 2+1 & $A_{\mu}$, $\phi$  \\
\hline
Bulk  & 9+1 & $C_{p-1}$ & 3+1 & $C_{\mu}$ \\
\hline
\end{tabular}
\end{center}
\caption{Correspondence between the field content
and dimensionalities for the two models.\label{toy_model}}
\end{table}

We choose the complex scalar to be an Abelian Higgs field instead of a
tachyon, since the tachyon field develops singularities in finite time
and the simulation can only run until the first defect forms. The Higgs
captures the essential features of the tachyon, but is well behaved
after the defects have formed, so we can follow the evolution. The
action of the model is:%
\footnote{%
    Here and throughout we use the $[{-}{+}{+}{+}]$ metric convention
    and geometrical normalization for gauge fields, so covariant
    derivatives are of form $D_\mu = \partial_\mu -\i A_\mu$.  The scalar
    field charge is chosen to be 1.  The mass dimensions
    of the parameters and fields appearing in the Lagrangian are,
    $[A]=[C]=1$, $[\phi]=1/2$, $[g_{\rm bulk}^2]=0$, $[g_{\rm
    brane}^2]=[\lambda]=1$, and $[c^2_{\rm cs}]=0$.
    }
\be
{\mathcal L} = \int_{{\mathcal M}_{3}} \!\!\!dx^2dt 
\left[-\frac{1}{4g_{\rm brane}^2}F^2 -
D_{\mu}\phi D^{\mu}\phi^{*}-V\left(\phi\right)\right] -
\frac{c_{cs}}{2}\int_{{\mathcal M}_{3}} \!\!\!F \wedge C +
\int_{{\mathcal M}_{4}} \!\!\!dx^3dt \left[-\frac{1}{4g_{\rm bulk}^2}H^2\right]
\,.
\label{Lagrangian}
\ee
We denote by $F=dA$ the field strength of the field $A$ and by $H=dC$
the field strength of the field $C$. The scalar covariant derivative
is $D_\mu = \partial_\mu -iA_\mu$.  One can easily check that the
action is invariant under gauge transformations for each one of the
fields $A$ and $C$.  We take the standard symmetry breaking form for the
scalar potential,
\be
V(\phi) = \lambda \left( \phi^* \phi - v^2/2 \right)^2 \, .
\ee

It is straightforward to derive the equations of motion which follow
from this action.  The
simplest one to derive is the equation of motion for the scalar field
since the scalar $\phi$ does not couple directly with the bulk field:
\be
\partial_{\mu}\partial^{\mu}\phi-2{\i}A^{\mu}\partial_{\mu}\phi-
A_{\mu}A^{\mu}\phi - {\i}\phi\partial^{\mu}A_{\mu}+
\frac{\partial{V}}{\partial{\phi^{*}}} = 0 \,.
\ee
One usually chooses the Lorentz gauge $\partial^{\mu}A_{\mu} = 0$ to further
simplify the equation above. This can be convenient when solving for a static,
isolated defect, but is not the most convenient one when performing
lattice simulations.  The equations of motion for the gauge fields are
the usual ones with an extra term coming from the Chern-Simons interaction:
\baray
\label{gauge_eqmo}
\partial_{\mu}F^{\mu\nu} +
{\i} g^2_{\rm brane} \left( \phi D^{\nu}\phi^{*}-\phi^{*}D^{\nu}\phi\right) +
c_{cs}g^2_{\rm brane} \frac{\epsilon^{\nu\alpha\beta}}{2}
H_{\alpha\beta} &=& 0\,,
\\
\label{H_eqmo}
\partial_{\mu}H^{\mu\nu} -
c_{cs} g^2_{\rm bulk}
\frac{\epsilon^{\alpha\beta\nu}}{2}
F_{\alpha\beta}\delta\left(z\right) &=& 0 \, .
\earay

\subsection{Vortex defects}
\label{Vortex_content}

At vanishing Chern-Simons interaction, $c_{cs}=0$, these equations
support a topological defect, the Nielsen-Olesen vortex
\cite{Nielsen:1973cs}.  In any nonsingular gauge, choosing cylindrical
coordinates $(\rho,\theta,z)$ centered on the vortex, the phase of the scalar
field $\phi$ winds by $2\pi n$, $n$ an integer, in going around the
vortex.  At large radii, the scalar rests in its broken phase minimum
$|\phi|^2=v^2/2$, and the gradient
energy which the phase change would give rise to
is exactly canceled by a gauge field, which in
a particularly convenient gauge takes the value,
\be
\sqrt{2} \phi(\vec r) = v e^{in\theta} \, , \qquad
A_\theta(\vec r) = \frac{n}{\rho} \, .
\ee
Close
to the core of the defect, the gauge field takes a smaller value and the
scalar field leaves the vacuum manifold, with $\phi=0$ at the exact
vortex center.  The defect core costs finite energy and carries a net
magnetic flux of $2\pi n$.  Far from the vortex, the fields are
locally in vacuum and all excitations are massive; so the mutual
interactions of two vortices or a vortex-antivortex pair are
exponentially suppressed with separation.

Examining \Eq{gauge_eqmo} shows why these conclusions will change
somewhat at nonzero values for the Chern-Simons coupling $c_{cs}$.
Working in the same gauge, we see that only $F_{\rho\theta}$ is nonzero, so
only $C^0$ is sourced by the Chern-Simons term.  Its equation of motion
is,
\be
\nabla^2 C^{0}-c_{cs} g^2_{\rm bulk} F_{12}\delta\left(z\right) = 0 \, .
\label{C_eqmo}
\ee
The component $F_{12}=F_{\rho\phi}$ acts as a surface charge density
sourcing the bulk field.

Our goal is to try to understand the structure and energetics of
an isolated defect, or a widely separated defect anti-defect pair, when
this Chern-Simons interaction is taken into account.  As a byproduct, we
will determine the force between defects, which will no longer fall off
exponentially.  The force will be attractive between defects of opposite
winding number, but repulsive between defects of the same winding
number; our toy model does not include the dilaton and the graviton
fields, which can provide compensating attractive interaction, so we cannot
obtain the BPS states that correspond to two identical, parallel, branes.

One can perform a Kaluza-Klein reduction of the 3+1
dimensional bulk theory down to 2+1 dimensions.  This will yield a
theory containing a light $\phi$ and $A$ field and a massless $C$ field,
together with a Kaluza-Klein tower of heavier $C$ field modes.  It is
possible to solve for the defect structure using this effective theory.
Since the brane fields are directly the fields appearing in the 2+1D
effective theory, their equations of motion change the least.
The equation of motion for $\phi$ is unchanged, and the expression for
$A^\mu$, \Eq{gauge_eqmo}, is changed only by the replacing $H$ with a
sum over KK states of $H$ from each KK state.  The change for the bulk
field $C$ is slightly more complicated.
The KK decomposition is a Fourier series expansion,
so $\delta(z)$ in $z$ space becomes the constant $1/2\pi R$
in each KK mode.  Therefore, the equation of motion for the $C^0$ field,
\Eq{C_eqmo}, becomes,
\be
\left(\nabla^2 - M_{\rm kk}^2\right) C_m^0 
= \frac{c_{\rm cs} g^2_{\rm bulk}}{2\pi R}
 F_{12} \, , \qquad M_{\rm kk}^2 = \left( \frac{m}{R}\right)^2 ,
\ee
with $m$ the index of the KK field (the KK wavenumber).  

\begin{figure}
\centerbox{0.7}{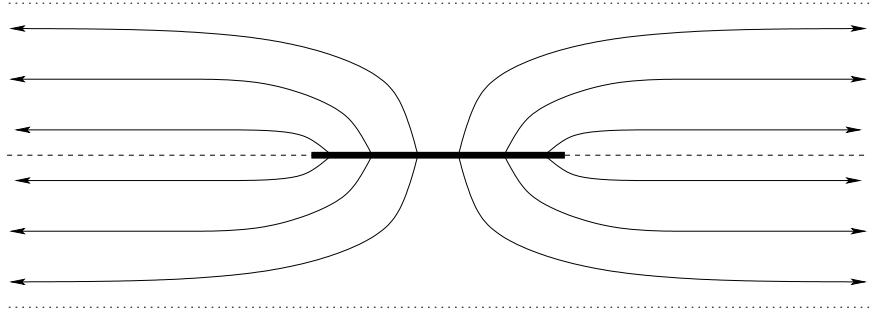}
\caption{\label{fig_cartoon} Cartoon of how the bulk $H$ electric field,
sourced by the $F$ flux at the defect core, spreads out across the
extra dimension.  The dashed line is the brane, the thick bar is the $F$
field strength in the defect core, the lines with arrows are the
$H=\nabla C^0$ electric flux lines, and the dotted line is the periodic
boundary in the bulk direction.}
\end{figure}

We will not attempt to solve these equations in detail near the
defect core, but will content ourselves with an understanding of the
defect appearance at radii larger than the symmetry breaking scale,
$\rho \gg 1/g_{\rm brane}v$, and larger than the radius of the extra
dimension, $\rho \gg R$.  In this case, the appearance of $M_{\rm kk}^2$
in the equation for $C^0_m$
leads to an exponential decay of $C^0$ away from the defect core for all
KK states but the zero mode, which instead falls according to the 2D
Coulomb's law:
\baray
\label{eq_C}
C^0(\rho) & = & \frac{c_{\rm cs} g^2_{\rm bulk}}{2\pi R} \frac{\Phi}{2\pi}
\ln(\rho/\rho_0) \, , \\
\Phi & \equiv & \int d\theta \rho d\rho F_{12} = \oint \rho d\theta
A_\theta \, ,
\earay
where $\Phi$ represents the total magnetic flux associated with the brane
defect, and $\rho_0$ is a constant of integration.  Physically, the
charge distribution for the $C$ field is a disk lying on the brane at
the defect core, but the $H$ electric field spreads out from this disk
and becomes uniform across the extra dimension at a distance of a few
$R$, as illustrated by the cartoon in figure \ref{fig_cartoon}.

Unlike the case of vanishing $c_{\rm cs}$, the magnetic flux $\Phi$ will
not equal $2\pi$.  There are two ways to see this; by
considering the equations of motion, or by considering energetics.
Looking first at the field equations, the requirement that the scalar
field phase change by $2\pi n$ around the defect is topological
and is not modified by the Chern-Simons term.  However, the presence of
the Chern-Simons term in \Eq{gauge_eqmo} means that either the field
strength or the Higgs field current must now be nonzero.  The defect
core is defined as the region where the gauge field strength is
non-negligible; outside the core we expect $F_{12}=0$, so only the Higgs
field gradient term can balance the Chern-Simons term.  This will happen
if the gauge field is smaller than $A_\theta = n/\rho$ (its value in the
Nielsen-Olesen solution), corresponding to a flux which is smaller
than $\Phi = 2\pi n$, because in this case the derivative and gauge
field parts of the covariant derivative $D_\theta$ will not cancel when
acting on $\phi = e^{in\theta} v/\sqrt{2}$.  In detail, the equation of
motion becomes,
\baray
\i g_{\rm brane}^2 \left( \phi \frac{1}{\rho}D_\theta \phi^* 
- \phi^* \frac{1}{\rho} D_\theta \phi \right)
& = & c_{\rm cs} g^2_{\rm brane} \partial_\rho C^0 \, ,
\\
\frac{v^2}{\rho} \left(n-A_\theta/\rho \right)
& = & c_{\rm cs} \frac{c_{\rm cs} g^2_{\rm bulk}}{2\pi R}
\frac{\Phi}{2\pi} \frac{1}{\rho} \, ,
\earay
where we used \Eq{eq_C} in the second line.  Solving for $\Phi$,
remembering that $\Phi=2\pi A_\theta \rho$, we find,
\baray
v^2 \left( n - \frac{\Phi}{2\pi} \right) & = & 
\left( \frac{c_{\rm cs}^2 g_{\rm bulk}^2}{2\pi R} \right)
\frac{\Phi}{2\pi} \, , \\
\Phi & = & 2\pi n \left[ 1 + \frac{c_{\rm cs}^2 g_{\rm bulk}^2}
 {2\pi R v^2} \right]^{-1} \, .
\earay

\subsection{Defect Energetics}

Alternatively, we can understand this result by considering energetics.
If the gauge field flux does not equal $2\pi n$, there will be a
nonvanishing Higgs field gradient energy.  However, the larger the flux,
the larger the $C$ electric field.  The vortex is the minimal energy
configuration subject to the topolocial condition on the phase of
$\phi$.  Therefore, the gauge field will take the value
which minimizes the sum of these two energy densities;
\be
\frac{d}{dA_\theta} \left( D_\theta \phi^* D_\theta \phi
+ \frac{2\pi R}{2g_{\rm bulk}^2} \partial_\rho C^0 \partial_\rho C^0
\right) = 0 \, .
\ee
In the first term, $D_\theta \phi = (\i n/\rho - \i A_\theta) \phi$, and
$\phi^*\phi = v^2/2$ (since otherwise there would be an extensive
potential energy).
In the second term, $C^0$ is determined from \Eq{eq_C}.  Also, $A_\theta
= \Phi/2\pi\rho$, as previously.  The
minimization therefore requires,
\be
\frac{d}{d\Phi} \left[ \frac{v^2}{2\rho^2} \left( n-\frac{\Phi}{2\pi}
  \right)^2 + \frac{2\pi R}{2g_{\rm bulk}^2} 
  \left(\frac{c_{\rm cs}g_{\rm bulk}^2 \Phi}
  {(2\pi)^2 R \rho} \right)^2 \right] = 0
\quad \Rightarrow \quad
\Phi = 2\pi n \left[ 1 + \frac{c_{\rm cs}^2 g_{\rm bulk}^2}
 {2\pi R v^2} \right]^{-1} \, ,
\ee
the same answer we obtained directly from the equations of motion.

This calculation also allows us to find the energy density far from the
brane core, which is,
\be
\epsilon = | D_\theta \phi |^2 
+ \frac{2\pi R}{2 g_{\rm bulk}^2}|\nabla C^0|^2 \, ,
\ee
which using the above result for $\Phi$ equals,
\be
\epsilon = \frac{v^2 n^2}{2\rho^2} \frac{a}{1+a} \, , \qquad
a \equiv \frac{c_{\rm cs}^2 g_{\rm bulk}^2}{2\pi R v^2} \, .
\ee
Note several features of this energy density.  First and most important,
it falls off as $1/\rho^2$ as one increases the distance $\rho$ from the
core of the defect.  Second, as the Chern-Simons coupling $c_{\rm cs}$ is
changed from 0 to $\infty$, the energy density far from the core changes
smoothly from 0, the value for a ``local'' (Abelian Higgs) topological
defect, to $v^2 n^2/2\rho^2$, the value for a ``global'' topological
defect (one for a complex, gauge singlet scalar field).  
%
The total energy of a defect, $\int d\theta \rho d\rho \epsilon$, is
logarithmically divergent;
\be
E = \int d\theta \; \rho d\rho \; \frac{v^2 n^2}{2\rho^2}
\frac{a}{1+a} = \pi v^2 n^2 \frac{a}{1+a} \ln \rho_{\rm max}/
\rho_{\rm min} \, .
\label{eq:tot_energy}
\ee
Here $\rho_{\rm min} \sim 1/g_{\rm brane} v$ is a constant of
integration which also absorbs the finite contribution from the defect
core, and
$\rho_{\rm max}$ is the distance where interactions with the fields of
other defects become important.  For instance, for a defect-antidefect
pair of separation $r$, the energy is $E = 2\pi v^2 n^2 (a/[1{+}a])
\ln r/\rho_{\rm min}$.  This behavior shows that a defect antidefect
pair will feel an attractive $1/r$ interaction, which is proportional to
$c_{\rm cs}^2$ when the Chern-Simons coupling is small but has a finite
limit when it is large.  The Chern-Simons term has replaced the
exponentially weak long-range interaction with one more typical of
Coulomb interactions in two dimensions.

One feature of \Eq{eq:tot_energy} is that the energy associated with a defect
is never larger than the energy of a global defect, that is, a defect
without either the brane field $A$ or the bulk field $C$.  We have shown
this for the large distance contribution, but it is also true in the
core; for while the bulk field $C$ costs energy, it only does so if
the gauge field $A$ takes on a nonvanishing flux.  The energetics are
that the flux is present because it reduces the scalar field gradients,
and it will always take a value which reduces the scalar gradients more
than it increases field strength energies.  Therefore the energetics of
defect formation are at least as favorable as in a global defect model.
Also note that the {\em larger} the extra bulk dimension, the 
{\em smaller} the parameter $a$ and therefore the energy, as the $C^0$
field strength spreads out over a larger dimension and therefore carries
a smaller field strength squared.

We can now understand why a network of defects is energetically
permitted to form in the process of brane-antibrane annihilation, which
is assumed to start with the tachyon field near zero.  The analogous
situation in the toy model is that the scalar field starts at zero,
though with small quantum fluctuations away from zero.
Consider an isolated defect, hypothetically formed after
brane-anti-brane annihilation.  There is a finite contribution to its
energy from the defect core region, and a
contribution arising from large $\rho$, which grows like
$\log \rho$. However, inside the same radius, the potential energy
initially available from the
the scalar field grows like the area, $\rho^2$, and this
always wins over the $\log \rho$ dependence. We can therefore conclude
that for a large enough radius, $\rho$, there is always enough energy
available for creating the defect. The reason is that the energy of
the final defect is coming from the potential energy of the brane
scalar field, or the tachyon in the String Theory case, and not from
the quantum deSitter fluctuations.

This study has clarified the behavior of defects in the toy model and
argued that their formation is not obstructed by energetic
considerations.  We now will verify these conclusions, and study the
dynamics of the formation and evolution of the system of vortices which
forms in brane-antibrane annihilation in this toy model, by studying it
via nonperturbative lattice techniques.

\section{Lattice Simulation}

The previous section established a toy model and made preliminary
investigations of the defect structure and energetics in this model.
What we really want is to study the dynamics of this model and its
defect formation in a nonperturbative way.  Therefore we have
implemented the toy model on the lattice.
Excluding the Chern-Simons coupling, which will be presented in
an appendix, the lattice regularization of the brane and bulk action
is the usual one for an Abelian Higgs model in $2+1$ dimensions and a
Abelian gauge field in $3+1$ dimensions. The fact that the fields are
Abelian allows us to use the non-compact formulation of the lattice action.

The lattice equations of motion are most easily evolved in temporal
gauge,  $A^{0}=0$ and $C^{0} = 0$.  This choice is immaterial so long
as we concentrate on gauge invariant observables.  It makes it slightly
harder to verify the field configuration for the vortex discussed in the
previous section, since the two configurations are related by a
time-dependent gauge transformation.  However, we can still
compare lattice studies to our results there using only gauge invariant
quantities such as field strengths.  In particular, defects are easily
identified by the peak of magnetic field strength at their core, along
with the peak in energy density.

Besides studying the static vortex solutions, the lattice implementation
allows us to study the dynamical
evolution of the fields, including
the formation and annihilation of the vortices and
the flow of energy from the brane into the bulk.

The goal of the simulations is to see whether defects form, and how the
defect network subsequently evolves if they do form, when the fields are
prepared in a state designed to mimic the initial state of the tachyon
field right after a brane collision.  Namely, we begin with vanishing
gauge fields $A,B,\partial_t A,\partial_t B$, and with the scalar field
$\phi$ at the unstable symmetric point $\phi=0$.  To simulate initial
vacuum fluctuations, we add small fluctuations 
in the scalar field.  These will grow unstably, triggering the phase
transition to the state where $\phi^*\phi \simeq v^2/2$ almost
everywhere.  We look for defects as localized bundles of $F_{12}$ field
strength and $|D \phi|^2$ gradient energy, and explore their subsequent
evolution.  Our lattice simulations are generally performed using a
lattice spacing $av^2=0.5$, with couplings $\lambda=g_{\rm brane}^2=v^2$
and $g_{\rm bulk}^2=1$.

We find that defects do indeed form.  At vanishing Chern-Simons number,
the defects' mutual attraction is exponentially weak, and they persist
almost indefinitely in a large lattice volume.  (We use square, periodic
lattices.  Because of the periodicity of the lattice, the net winding
number of all defects, $\sum n$, is always zero, that is, there are
always equal numbers of defects and anti-defects, so every defect always
has a partner which it can annihilate off against.)  At finite
Chern-Simons number, there is long range attraction and the defects are
observed to annihilate off much faster, though the larger the lattice
volume, the longer it takes.

In order to confirm the two effects discussed earlier, we perform
simulations in which we change each of two ``control'' parameters,
the size of the extra dimension and
the strength of the Chern-Simons coupling. This way we can
see the effect of the size of the extra dimension on the energy of the
bulk field and the effect of the strength of the Chern-Simons coupling
on the magnetic flux through the core of the vortex.

\begin{figure}[tbp]
  \begin{center}
    \includegraphics[width=0.4\textwidth,angle=0]{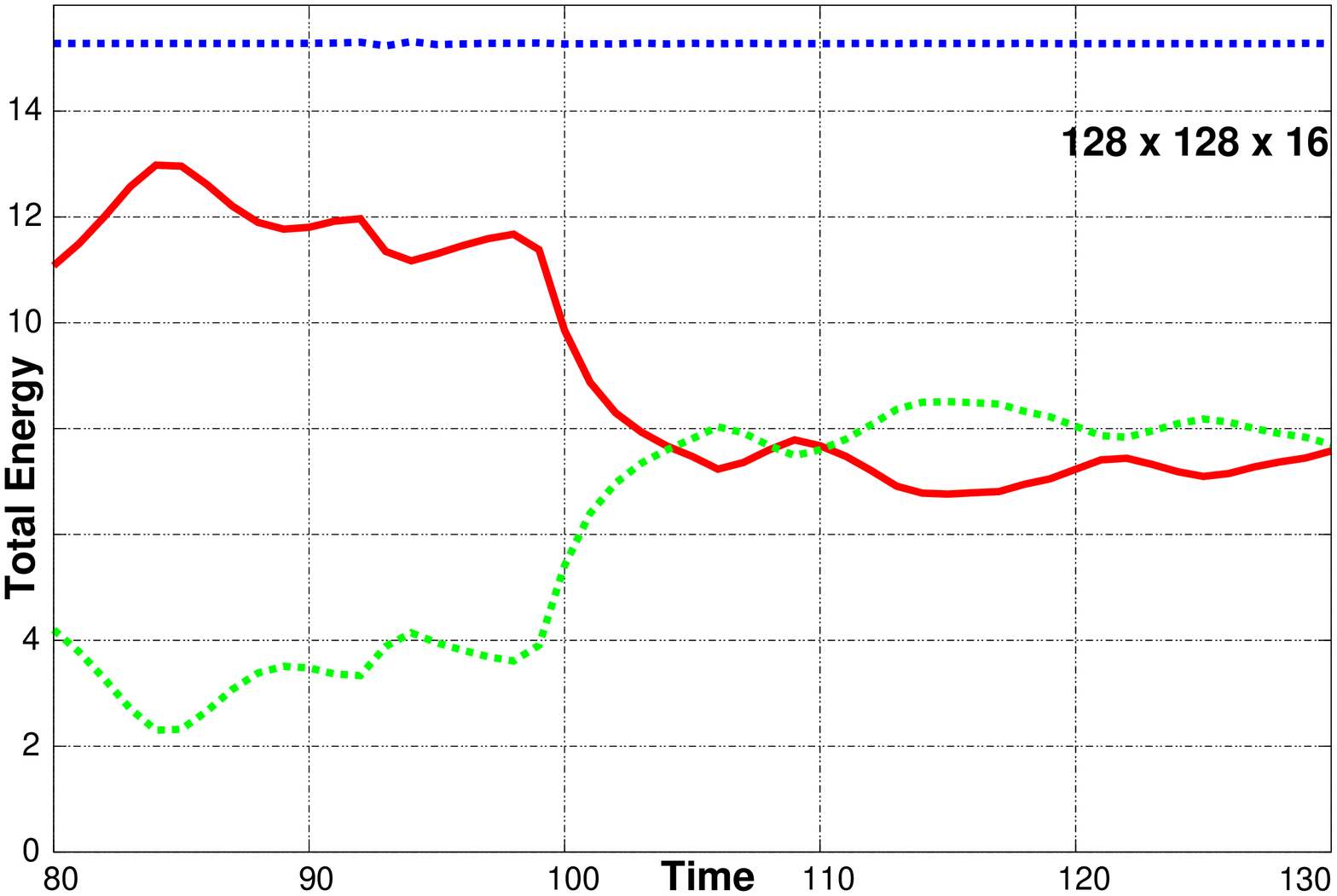}
    \hspace{6pt}	
    \includegraphics[width=0.4\textwidth,angle=0]{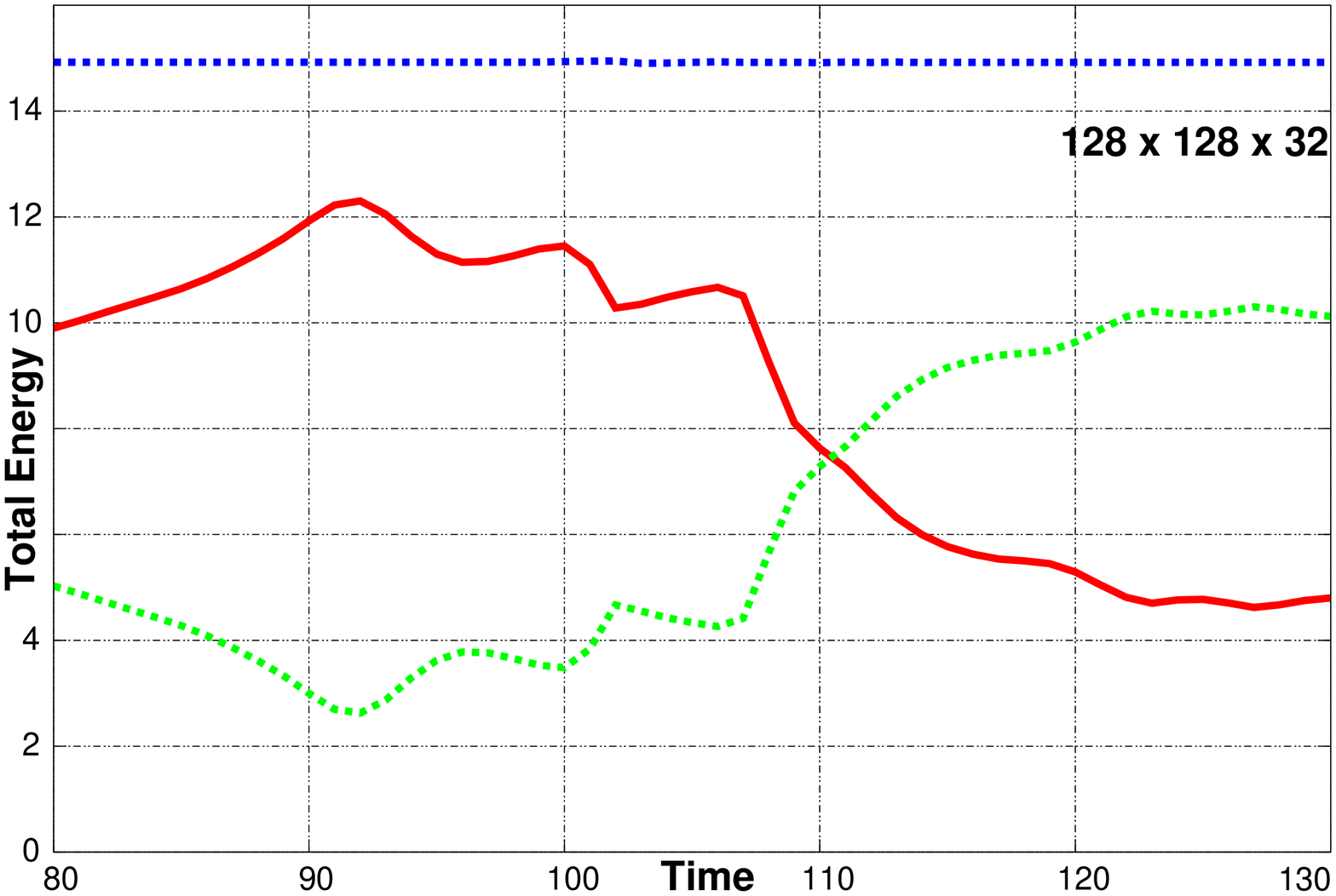}

    \vspace{12pt}

    \includegraphics[width=0.4\textwidth,angle=0]{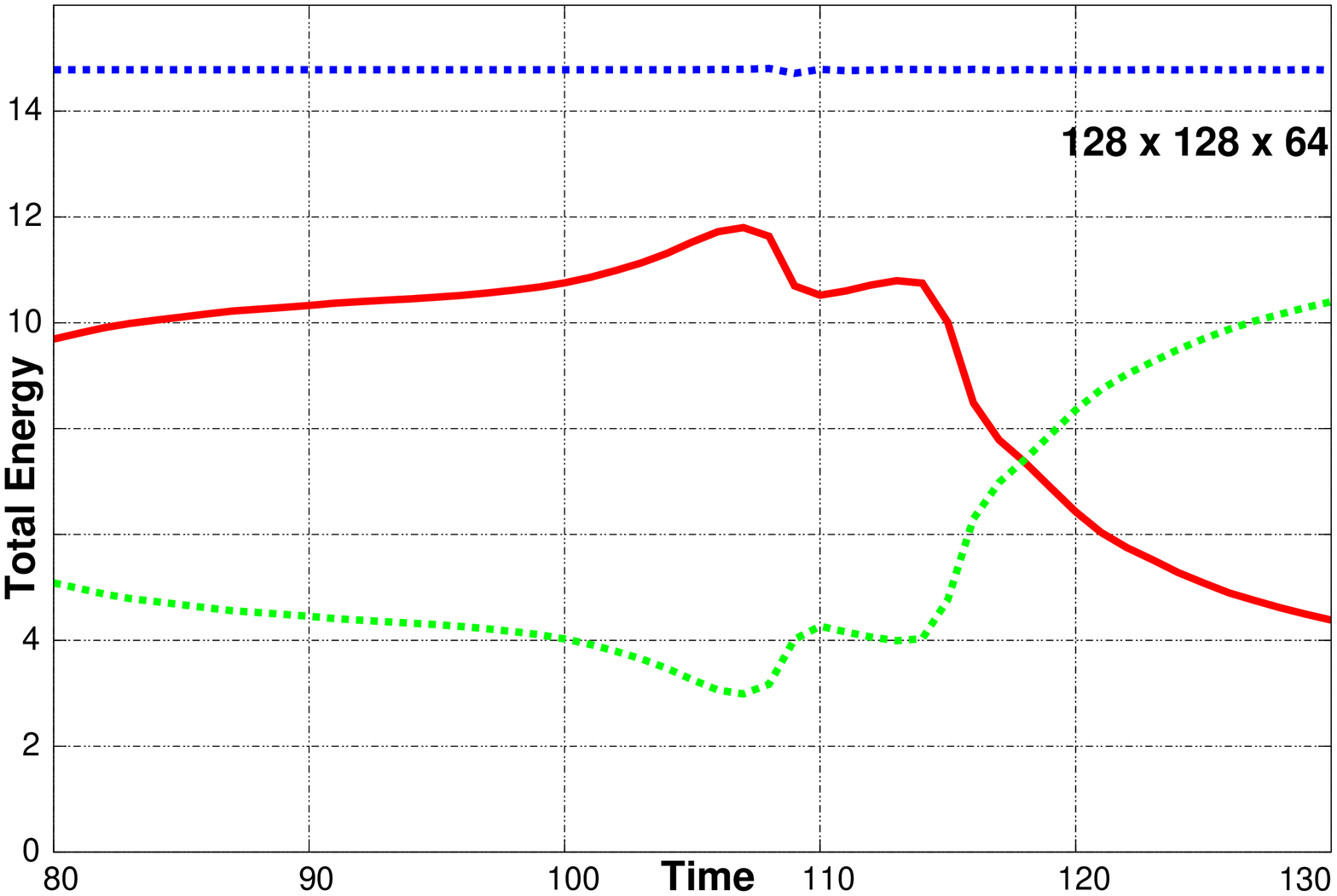}
    \hspace{6pt}
    \includegraphics[width=0.4\textwidth,angle=0]{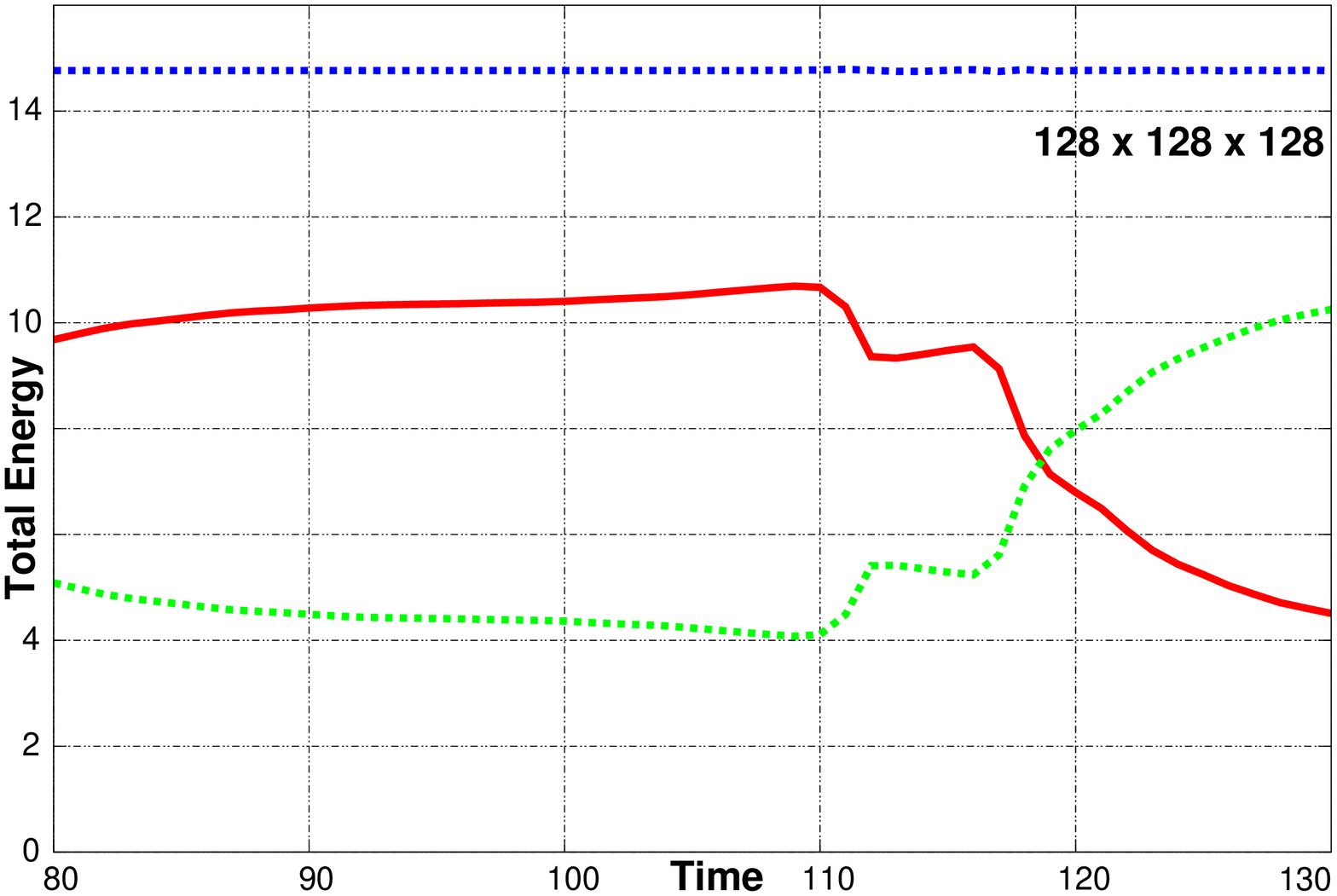}
  \end{center}
  \caption{(Color online)
  Energy in bulk and brane fields as the bulk size is varied.
  Each simulation begins with a well separated brane-antibrane pair.
  The brane energy is the solid (red) curve and the bulk energy is the
  dotted (green) curve. The abrupt discontinuity is a brane-antibrane
  annihilation event, which occurs later and later as the extra
  dimension is made larger, since the interaction becomes weaker.
  \label{fig2}}
\end{figure}

The effect of $R_{\rm bulk}$ is illustrated in Figure \ref{fig2}, which
shows the evolution of the energy density on the brane and in the bulk
for a configuration which was initially evolved with strong Hubble
damping until there is a single defect-antidefect pair, and then allowed
to evolve freely.  The total energy of the defect pair grows smaller as
the extra dimension becomes larger.
This is expected since a smaller extra dimension imposes larger
gradients of the bulk field.
We also observe that the defects annihilate more
slowly if the extra dimension is larger, consistent with our analysis of
the energetics.  (The larger extra dimension lets the bulk field spread
out more, weakening the Coulomb attraction of the pair.)  For the
largest extra dimensions we explore (64 and 128 across), the defect pair
start close enough together that the Coulombic interaction is
effectively 3D rather than 2D.

As the two defects approach each other there is an increase
in the energy of the brane fields and a decrease of the one in
the bulk field. The decrease in the bulk field energy is due to the
fact that the pair of charges at small separation will generate a
dipole field at large distances and this field has a lower total
energy than two widely separated charges. The increase in the
brane field energy can be thought of as the energy of motion of the two
defects plus radiated brane field energy as they interact and
annihilate. 

\begin{figure}[tbp]
\begin{center}
\includegraphics[width=0.8\textwidth,angle=0]{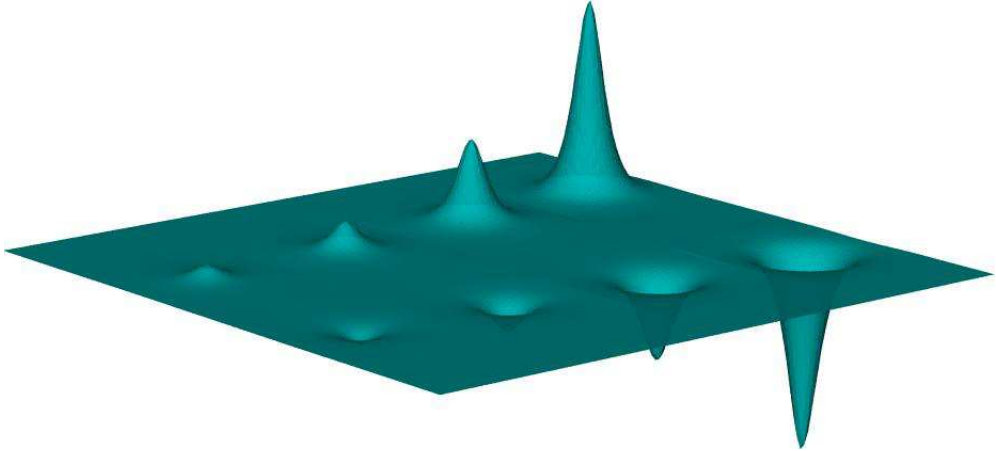}
\end{center}
\caption{ (color online)
The magnetic field strength distribution of a
defect-antidefect pair.  The figure shows the field strength
distribution for a series of defect-antidefect pairs 
in theories with progressively smaller
Chern-Simons interactions.  For the leftmost pair, the Chern-Simons term
almost eliminates the field strength; for the rightmost pair, the
Chern-Simons coupling has been turned off and the full field strength is
obtained. \label{fig3}}
\end{figure}

Changing the strength of the Chern-Simons coupling tells another story,
see Figure \ref{fig3}. Looking at the defects themselves, we can
see the back-reaction of the bulk field on the brane fields.
Increasing the strength of the coupling results in a
reduction of the magnetic flux through the core of the defect, in
agreement with the considerations of the previous section.

We also look at the energy transfer rate between the brane and the
bulk. One of the complications of a symmetry breaking transition such as
the one we are studying is, that the extra energy density associated
with the initial field potential energy has to ``go somewhere'' during
the simulation.  Frequently people carry out such simulations using
dissipative dynamics, meaning the extra energy is eaten up by friction.
This is justified in our case if the Hubble expansion rate is large.
Otherwise, for conservative evolution, the energy density typically
finds its way into the plentiful short wavelength excitations, where it
eventually becomes equipartitioned.  Because there is a brane-bulk
interaction in our simulations, the bulk
acts as an energy sink which can be much more efficient
than Hubble expansion. If the energy of the brane is drained quickly,
the fields will have less time to homogenize, increasing the density
of topological defects.
The question of energy leakage is also relevant when studying reheating.

\begin{figure}[tbp]
  \begin{center}
    \includegraphics[width=0.45\textwidth,angle=0]{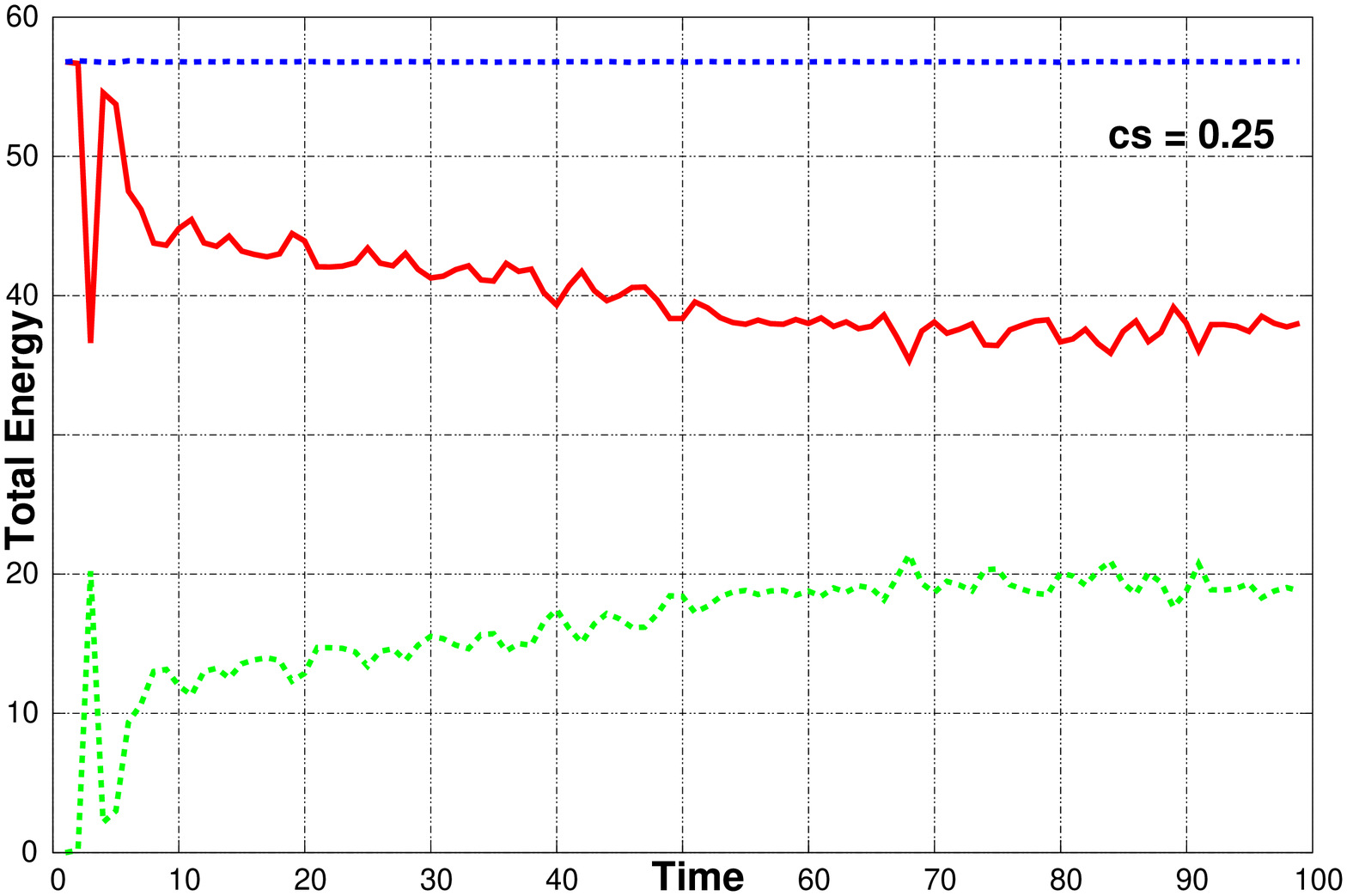}
    \hspace{6pt}	
    \includegraphics[width=0.45\textwidth,angle=0]{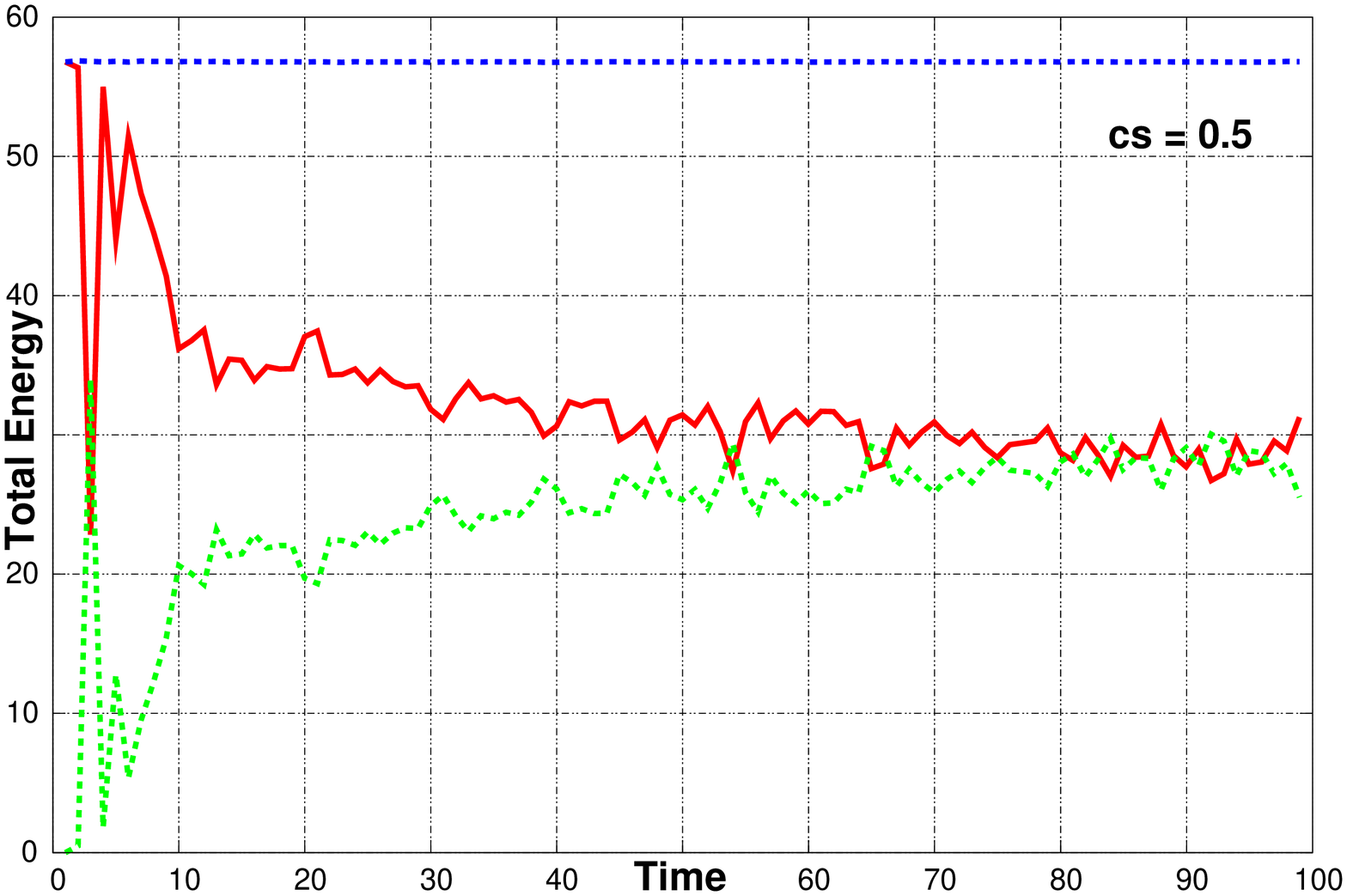}

    \vspace{12pt}

    \includegraphics[width=0.45\textwidth,angle=0]{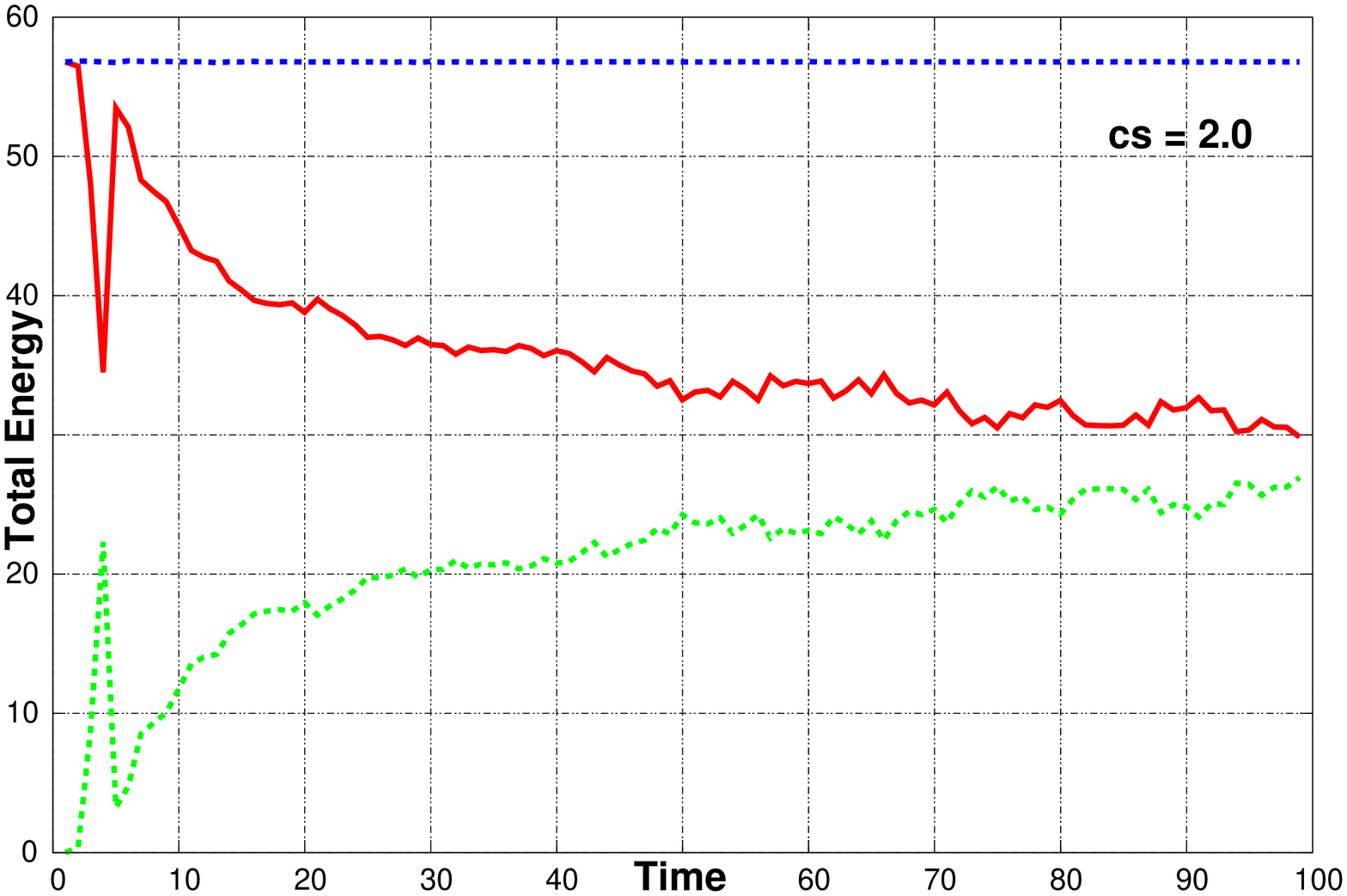}
    \hspace{6pt}
    \includegraphics[width=0.45\textwidth,angle=0]{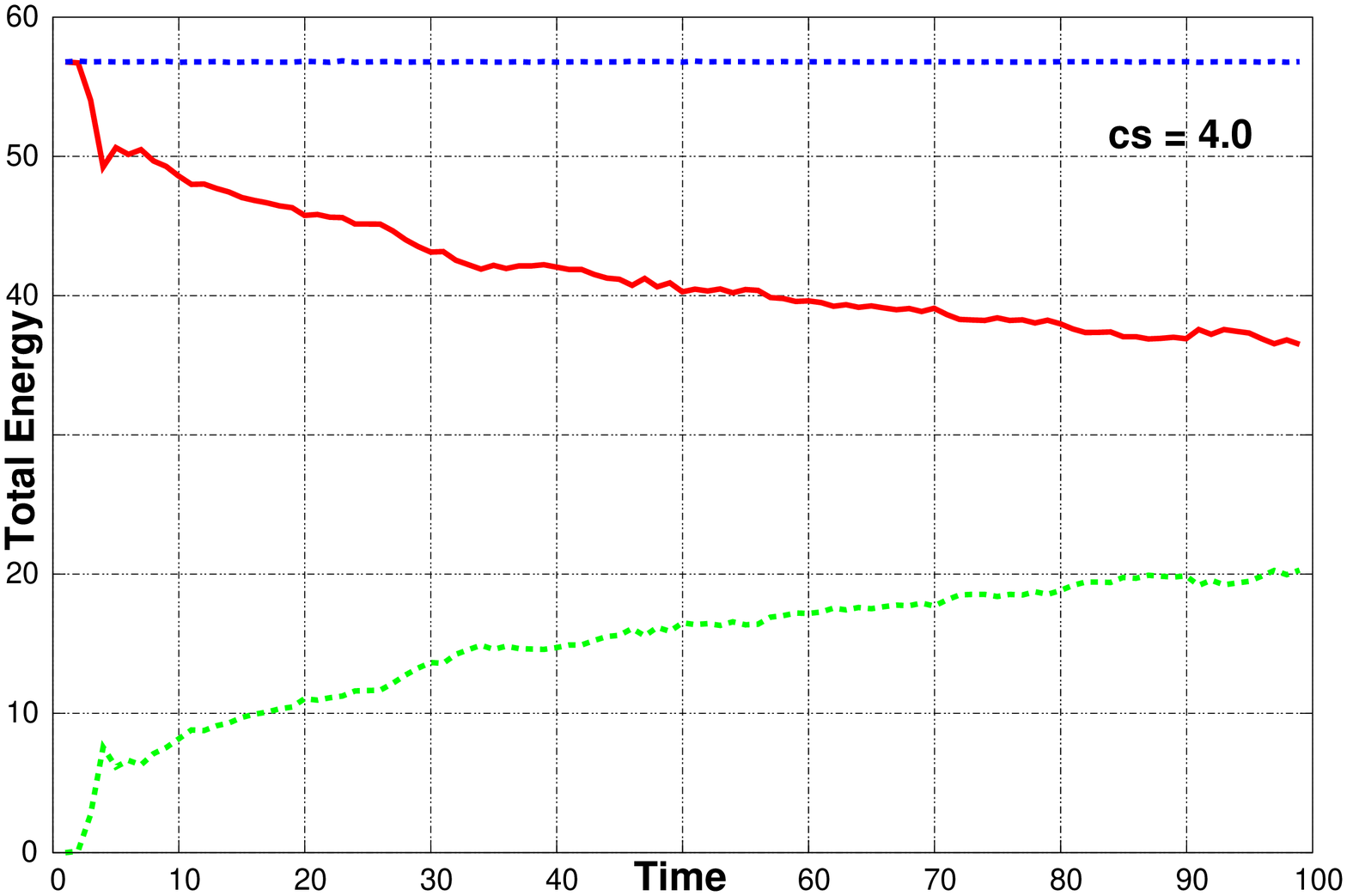}
  \end{center}
  \caption{ (color online)
Energy transfer between the brane (solid red line) and the bulk (dotted
green line) for different
values of the Chern-Simons coupling. For large or small coupling
the transfer is slow, so the bulk does not act as an efficient
energy sink. A slow flow of energy can allow the brane scalar
field to homogenize, reducing the final density of defects.
\label{fig4}}
\end{figure}

When studying this effect we use a small lattice
$32\times 32 \times 8$ which allows us to evolve
the system for a much longer time.
As expected (see Figure \ref{fig4}), for a weak coupling the energy is
transferred slowly to the bulk, so the brane fields
have more time available to homogenize. However as the strength of the
coupling is increased the energy transfer rate reaches a maximum, and
then it decreases again. We can understand this effect if we take into
account the fact that the magnetic flux of the brane gauge field acts
as the electric charge for the bulk gauge field. We saw before that
when increasing the strength of the coupling the charge increases.
However, when the charge becomes too large, it becomes energetically
expensive for brane magnetic fields to appear at all; therefore a very
large Chern-Simons interaction simply prevents brane magnetic fields from taking large values,
turning off the mechanism available for exchange of energy between brane
and bulk.

\section{Discussion and Conclusions}

Previous studies of defect formation during tachyon condensation have
neglected the role played by bulk fields.  It was argued
\cite{Dvali:2003zj} that when bulk fields are included, the formation of 
topological defects is suppressed or prevented. The argument was that
the energy of the deSitter fluctuations is of the order
the Hubble constant during inflation, while the gradients of the fields
along the compact dimensions have corresponding energies $1/R \gg H$; so
there is not enough energy available for defect formation.

We have constructed a toy model that captures the essential features
of the tachyon condensation process when both the brane worldvolume
gauge field and the bulk RR field are included. In this model the
gradients of the bulk field are formed as a result of classical field
evolution and the energy is provided by the tension of the
brane-anti-brane pair. Since this energy can also be described as the
energy of the tachyon field placed at the top of the potential,
in the toy model this corresponds to the potential energy of an
Abelian Higgs field. We argued that inside a large enough area around a
defect there is always enough energy in the form of potential energy of
the scalar field such that there is no energetic constraint on the
formation of topological defects.

In terms of energetics, the presence of the bulk with a Chern-Simons
coupling increases the energy cost of a defect.  The increase is
greatest if the Chern-Simons coupling is large and if the extra
dimension is {\em small}.  However, the energy associated with a defect
never exceeds the value in the global defect model, that is, the value
it would take if there were no gauge field or if $g_{\rm brane}^2=0$.
The bulk field also changes the long range interactions between defects,
so that their mutual attraction or repulsion shows power law, rather
than exponential, dependence on separation.  This speeds up the
evolution of the defect network, so that defect-antidefect pairs find
each other and annihilate more efficiently.  Again, the defect network
evolution is never faster than in the global defect model, which is
known to support a defect network after a symmetry breaking transition.

Another effect we observe in the lattice simulations of the toy
model is the decrease of the magnetic flux unit through the vortex
core as a function of the Chern-Simons coupling constant. This flux
also acts as the charge density for the bulk field and if the
Chern-Simons coupling becomes too large this charge density becomes
very small, reducing the energy leakage from the brane to the bulk.

One drawback of the toy model comes from using an Abelian Higgs field
instead of a tachyon on the brane. We did this so that we could follow the
evolution of the defects after they have formed while the tachyon
develops singularities. At the same time the brane fields do not drop
out of the dynamics \cite{CarollianContraction} once the defects have formed.

\section{Acknowledgements}

We would like to thank A. Vilenkin and J. J. Blanco-Pillado for collaboration during the 
early stages of this work. This work is supported by grants from NSERC of
Canada and FQRNT of Qu\'ebec.

\appendix

\section{Leading order Chern-Simons coupling}
\label{appA}

Here we expand the Chern-Simons coupling of the brane and bulk fields to
second order in $\alpha^{\prime}$. The full expression is
\be
S^{D\overline{D}}_{RR}=T_{D9}\int C\wedge \Str e^{2\pi i
  \alpha^{\prime}{\mathcal F}} \,,
\ee
where
\be
C =\sum_{p=\text{odd}}\frac{\left(-{\i}\right)^{\frac{9-p}{2}}}
   {\left(p+1\right)!} 
C_{\mu_{0}\dots\mu_{p}}dx^{\mu_{0}}\wedge\dots\wedge dx^{\mu_{p}} \,.
\ee
The supertrace of the matrix is defined as,
\be
\Str M = \Tr \left(\begin{array}{cc}1&0\\0&-1\end{array}\right)M \,,
\ee
and ${\mathcal F}$ is the curvature of the superconnection, given by
\be
i{\mathcal F} = \left(\begin{array}{cc}
iF^{+}-T\overline{T}&D\overline{T}\\
DT&iF^{-}-T\overline{T}
\end{array}\right) \,.
\ee

Here we are interested in the particular case of codimension 2 defects, so the relevant coupling
will be with the $C_{p-1}$ RR-field that couples with the defects. We want to keep both the
tachyon and the gauge fields non-zero, so we will expand the exponential inside the supertrace
in powers of $\alpha^{\prime}$:
\baray  \!\!\!
\Str e^{2\pi i \alpha^{\prime}{\mathcal F}} &=& 
\Tr\left(\begin{array}{cc}1&0\\0&-1\end{array}\right) +
\left(2\pi\alpha^{\prime}\right)
\Tr\left(\begin{array}{cc}1&0\\0&-1\end{array}\right)
\left(\begin{array}{cc}iF^{+}{-}T\overline{T}&D\overline{T}
\\DT&iF^{{-}}{-}T\overline{T}\end{array}\right)
\nonumber \\
&& + \frac{\left(2\pi\alpha^{\prime}\right)^2}{2}\Tr\left(\begin{array}{cc}1&0\\0&-1\end{array}\right)
\left(\begin{array}{cc}iF^{+}{-}T\overline{T}&D\overline{T}\\DT&iF^{-}{-}T\overline{T}\end{array}\right)
\wedge
\left(\begin{array}{cc}iF^{+}{-}T\overline{T}&D\overline{T}\\DT&iF^{-}{-}T\overline{T}\end{array}\right)
\nonumber \\ &&
+ \dots
\nonumber \\
& = & \left(2\pi i\alpha^{\prime}\right)\left[F^{+} - F^{-}\right] +
\label{SuperTrace}
 \\ && 
\left(2\pi \alpha^{\prime}\right)^2\left[D\overline{T}\wedge DT
-iT\overline{T}\left(F^{+} - F^{-}\right)
-\hf (F^{+}\wedge F^{+} - F^{-}\wedge F^{-})\right] + \dots \,. \quad
\nonumber
\earay
The terms of the type $F\wedge F$ couple with  $C_{p-3}$ and count only in the formation
of codimension 4 defects. Therefore, in the case of a $D9-\overline{D9}$ pair,
the important couplings between the brane and the bulk fields are
\be
T_{D9}\int -{\i}C_{8}\wedge\left\{ \left(2\pi i\alpha^{\prime}\right)\left[F^{+} - F^{-}\right] +
\left(2\pi \alpha^{\prime}\right)^2\left[D\overline{T}\wedge DT
-iT\overline{T}\left(F^{+} - F^{-}\right)\right]\right\} \,.
\ee
Since we are interested in the simplest model which involves such a
coupling, when we construct the toy model we keep only the interaction
that corresponds to the leading term in $\alpha^{\prime}$,
\be
\left(2\pi \alpha^{\prime}\right)T_{D9}\int C_{8}\wedge \left[F^{+} -
  F^{-}\right] \,.
\ee

\section{Higher Dimensional Generalizations}

We can extend the toy model we used to a higher dimensional one which 
will have fewer differences with respect to the full string-theory model. 
The most straightforward generalization would be to consider a 
$3+1$-dimensional brane and a $9+1$ dimensional bulk. The field content 
of the worldvolume theory would still be the Abelian Higgs model,
but the bulk field will now be a rank-2
antisymmetric tensor field, which has the 
appropriate rank to couple to the 2-dimensional world-volume of a string. 
The Abelian Higgs model in $3+1$ dimensions admits stable string-like 
defects which in our model will be charged under the bulk field. 
The Lagrangian of the model is:
\be
{\mathcal L} = \int_{{\mathcal M}_{4}} \!\!\!d^3 x\,dt 
\left[-\frac{1}{4g_{\rm brane}^2}F^2 -
D_{\mu}\phi D^{\mu}\phi^{*}-V\left(\phi\right)\right] -
\frac{c_{cs}}{2}\int_{{\mathcal M}_{4}} \!\!\!F \wedge C +
\int_{{\mathcal M}_{10}} \!\!\!d^9x\, dt \left[-\frac{1}{12g_{\rm bulk}^2}H^2\right]
\,,
\ee  
where as before we denote by $F=dA$ the field strength of the field 
$A$ and by $H=dC$ the field strength of the field $C$. The equations of motion 
now become:
\baray
\partial_{\mu}F^{\mu\nu} +
{\i} g^2_{\rm brane} \left( \phi D^{\nu}\phi^{*}-\phi^{*}D^{\nu}\phi\right) +
c_{cs}g^2_{\rm brane} \frac{\epsilon^{\nu\alpha\beta\gamma}}{2}
H_{\alpha\beta\gamma} &=& 0\,,
\\
\partial_{\mu}H^{\mu\nu\lambda} -
c_{cs} g^2_{\rm bulk}
\frac{\epsilon^{\alpha\beta\nu\lambda}}{2}
F_{\alpha\beta}\; \delta\left(z\right) &=& 0 \, .
\earay

If we now want to study how the solution for a Nielsen-Olesen string is modified 
by the presence of the bulk field, for the brane fields we make the usual ansatz 
for a string placed along the $z$-direction: far from the string the
scalar and gauge fields are
\be
\sqrt{2} \phi(\vec r) = v e^{in\theta} \, , \qquad
A_\theta(\vec r) = z \frac{n}{\rho} \, ,
\ee
with $z$ a rescaling of the $c_{\rm cs}=0$ case which remains to be
determined.
We observe that for the bulk field only the $C^{03}$ component is sourced by the 
Chern-Simons term. The entire analysis done for the lowest-dimensional toy model 
can be applied here with the only difference being the expression of the masses for the 
KK modes, wich will be highly dependent on the details of the compactification.
For the simplest case, a torus, the expression is:
\be
M_{kk}^2 = \sum_{i=1}^{6}\frac{m_{i}^2}{R_{i}^2} \,.
\ee
However, when studying the energetics of the defects, only the zero mode
gives a potentially log-divergent contribution, and
the details of the compactification are relevant only to an energy
density associated with the core.  Also, as in the toy model studied in
the main text, the (global) defect with the $\phi$ field varying by a
$2\pi$ phase around the defect core but both the $F$ and $H$ field
strengths vanishing gives an upper bound on the defect energetics.
As in the lowest-dimensional toy model we have:
\baray
C^{03}(\rho) & = & \frac{c_{\rm cs} g^2_{\rm bulk}}{V} \frac{\Phi}{2\pi}
\ln(\rho/\rho_0) \, , \\
\Phi & \equiv & \int d\theta \rho d\rho F_{12} = \oint \rho d\theta
A_\theta \, ,
\earay
with $V$ the (6-dimensional) volume of the compact manifold.
This leads to the same result for the unit of magnetic flux 
through the vortex core:
\be
\Phi  =  2\pi n \left[ 1 + \frac{c_{\rm cs}^2 g_{\rm bulk}^2}
 {V v^2} \right]^{-1} \, ,
\ee
and for the energetics of the defect; the string tension involves a log,
$\ln(\rho_{\rm max}/\rho_{\rm min})$, with $\rho_{\rm max}$ the
inter-string separation and $\rho_{\rm min}$ the string core size;
\be
T = \pi v^2 n^2 \frac{a}{1+a} \ln \rho_{\rm max}/\rho_{\rm min} \, , 
\quad
a \equiv \frac{c_{\rm cs}^2 g_{\rm bulk}^2}{V v^2} \, .
\ee

We could also have considered higher dimensional branes wrapped on cycles of 
the compact manifold, but again only the zero modes of the brane fields would 
participate in the formation of the defects. 

\section{10 D global defects in the original theory}

The easiest setup to study the formation of defects in the original theory is 
to consider the action for a brane-anti-brane pair for only the tachyon field in 
flat space-time, with all the other fields turned off, since the lower-dimensional 
branes are in fact vortices of the complex tachyon field. The action is simply:
\be
S = 2T_{D9}\int d^{10}X \,e^{-2\pi\alpha^{\prime}T\overline{T}}\left[
1+8\pi\alpha^{\prime}\ln\left(2\right)D^{\mu}\overline{T}D_{\mu}T\right]\,.
\ee
One can also consider the action for lower-dimensional brane-anti-brane pairs. 
The equation of motion derived from the acion above is:
\be
\partial_{\mu}\partial^{\mu}T-
2\pi\alpha^{\prime}\overline{T}\partial_{\mu}T\partial^{\mu}T+
\frac{T}{4\ln 2} = 0 \,.
\ee
As in the case of the real tachyon field, one can approximate the profile of the
field with a linear one as the vortex will form at the place where $T = 0$, and 
solve the resulting equation for the slope of the profile. The resulting defect 
formed in the decay of a $Dp-\overline{D}p$ pair is a $Dp-2$ brane.

   In order to undestand this we have to go back to the calculation of the space-time action 
calculated on linear tachyon profiles and estimate the action in the limit of infinite slope.  
The calculation was done in Ref. \cite{Kraus:2000nj} and we reproduce the important points here.
In the case of a $D9-\overline{D}9$ pair the space-time action for a
linear tachyon profile is:
\be
S\left(y^{I}\right)=2T_{D9}\int dX^{10} \,e^{-2\pi\alpha^{\prime}T\overline{T}} 
\prod_{I=1}^{2}F\left(\pi\alpha^{\prime}y^{I}\right)
\ee
where $T^{I} = u^{I}X^{I}/\sqrt{\alpha^{\prime}}$ and $y^{I}=\left(u^{I}\right)^2$. The function
$F$ has the expression:
\be
F\left(x\right) = \frac{4^{x}x\Gamma\left(x\right)^2}{2\Gamma\left(2x\right)}
\ee
and in the large argument limit it takes the form:
\be
F\left(x\right) \simeq \sqrt{\pi x} \,.
\ee
Calculating the action on the profile $y^{1}\rightarrow\infty$ and $y^{2}\rightarrow\infty$ 
the authors of Ref. \cite{Kraus:2000nj} obtain:
\baray
&& S\left(y^{I}\right) = 2T_{D9}\int dX^{10}
e^{-\frac{\pi}{2}\left[y^{1}\left(X^{1}\right)^2+y^{2}\left(X^{2}\right)^2\right]}
F\left(\pi\alpha^{\prime}y^{1}\right)F\left(\pi\alpha^{\prime}y^{2}\right)
\nonumber \\
&& =2T_{D9}\int dX^{8} \sqrt{\frac{2}{y^{1}}}\,\sqrt{\frac{2}{y^{2}}}
\,\sqrt{\pi^2\alpha^{\prime}y^{1}}\,\sqrt{\pi^2\alpha^{\prime}y^{2}}
\;\rightarrow\;
4\pi^2\alpha^{\prime}T_{D9}\int dX^{8} \,.
\earay
The result gives the correct tension for a $D7$ brane, 
$T_{D7} = \left(2\pi\sqrt{\alpha^{\prime}}\right)^2T_{D9}$. Regarding the RR charge of the vortex, 
we can estimate it by using the result Eq. (\ref{SuperTrace}) in the expression of the coupling 
between brane fields and the bulk RR fields,
\be
S_{RR} = T_{D9}\int C_{8}\wedge\, e^{-2\pi\alpha^{\prime}T\overline{T}}
\left(2\pi\alpha^{\prime}\right)^2dT\wedge d\overline{T} = 
\frac{\left(2\pi\alpha^{\prime}\right)^2}{\alpha^{\prime}}T_{D9}\int C_{8}\,,
\ee
which again reproduces the correct result for the $D7$ brane RR charge.
 
These results allow us to obtain an upper limit for the density of defects formed, 
based only on energetic considerations. The brane tension is equal to the mass per 
unit volume for the brane and the result 
$T_{Dp-2} = \left(2\pi\sqrt{\alpha^{\prime}}\right)^2T_{Dp}$ tells us that we can have 
at most one defect on each patch of area 
$\left(2\pi\sqrt{\alpha^{\prime}}\right)^2 = \left(2\pi l_{s}\right)^2$
where $l_{s}$ is the string length. This is a very large density and it shows that 
constraints other than the energetic ones are more important in determining the final density 
of defects. 

In a realistic model we expect that a very important role will be that of the other fields that 
we have neglected so far, namely the dilaton and the graviton. These two fields have 
universally attractive interactions and their presence allows for the existence of BPS states in which 
there is no interaction between identical, parallel, branes. We expect the presence of these fields
to also change the evolution of the resulting network of defects, since our toy model allows for repulsive
interactions between same-charge defects, while no repulsive interactions 
(except for very special situations, see \cite{Tye:StringProduction}) are possible in the full 
String Theory model. 

\section{Lattice implementation}

The implementation of gauge fields on the lattice was first developed by
Wilson \cite{Wilson74} and is by now standard.  To render the number of
degrees of freedom finite, the scalar field $\phi$ is taken only to
reside at a discrete set of points, a cubic (or square) lattice with
spacing $a$.  Since we will want a finite-timestep update, it is also
convenient to define the theory from the very beginning on a spacetime
lattice, with temporal lattice spacing $a_t \ll a$.
The gauge fields are a connection (rule for parallel
transport) and must be defined on the links (lines between nearest
neighbor lattice sites).  Therefore the lattice variable is
\be
\AL^\mu(x) = \int_x^{x+a \hat{\mu}} A\cdot dl \sim a A^\mu \, ,
\ee
with $\hat{\mu}$ the unit vector in the $\mu$ direction.
In a non-Abelian theory it is necessary to treat the exponent of this
variable, rather than $\AL$ itself, as the natural variable, but
in an Abelian theory we are free to consider either $\AL^\mu$
or $\exp(-i \AL^\mu)$ as the native variable, and we will choose
to use $\AL^\mu$ (the noncompact formulation).  Note that the
variable $\AL^\mu(x)$ is centered at $x+a\hat{\mu}/2$

The
terms in the action, \Eq{Lagrangian}, become,
\baray
\label{phi_latt}
\int_{{\cal M}_3} D_i \phi D^i \phi^* & \rightarrow & 
a^2 a_t \sum_{x\in {\cal M}_3} \sum_i
\frac{| e^{-\i \AL^i}\phi(x{+}a\hat{i})-\phi(x)|^2}
{a^2} \, , \\
\int_{{\cal M}_3} D_0 \phi D_0 \phi^* & \rightarrow & 
a^2 a_t \sum_{x\in {\cal M}_3}
\frac{| e^{-\i \AL^0}\phi(x{+}a_t \hat{t})-\phi(x)|^2}
{a_t^2} \, , \\
\int_{{\cal M}_3} F_{ij} F^{ij} & \rightarrow & 
a^2 a_t \sum_{x\in {\cal M}_3}
\sum_{ij} \frac{[\AL^i(x) + \AL^j(x{+}a\hat{i})
  - \AL^i(x{+}a\hat{j}) - \AL^j(x)]^2}{a^4}
\, , 
\earay
and similarly for $F_{0i} F^{0i}$ and the $H^2$ terms.  These terms are
all very standard in the lattice community.  For the Chern-Simons term,
the magnetic field is the sum of $A$ fields going around a square,
\be
\label{FL_is}
\FL^{ij} = \frac{1}{a^2} \left( \AL^i(x)-\AL^i(x{+}a\hat{j})-\AL^j(x)
+\AL^j(x{+}a\hat{i}) \right) = 
\begin{picture}(80,20)
\thicklines
\put(20,-10){\line(1,0){40}}
\put(60,-10){\line(1,1){20}}
\put(80,10){\line(-1,0){40}}
\put(40,10){\line(-1,-1){20}}
\put(20,-10){\vector(1,0){24}}
\put(60,-10){\vector(1,1){12}}
\put(80,10){\vector(-1,0){24}}
\put(40,10){\vector(-1,-1){12}}
\end{picture} \,,
\ee
which, note, is centered at $x+(\hat{i}{+}\hat{j})a/2$.  Therefore the $C$
field must be averaged over the four sites, $x$, $x{+}a\hat{i}$,
$x{+}a\hat{j}$, and $x+a(\hat{i}{+}\hat{j})$.  Further, since $C^0$
lives at the half time-step, it must be averaged over $C(t)$ and
$C(t-a_t/2)$;
\be
\label{latt_CS}
\int_{{\cal M}_3} F^{12} C^0 \rightarrow
a^2 a_t \sum_{x\in{\cal M}_3}
\FL^{12} \times
\frac{1}{8a_t} \sum \CL^0 =
\begin{picture}(90,20)
\thinlines
\put(20,-10){\line(1,0){40}}
\put(60,-10){\line(1,1){20}}
\put(80,10){\line(-1,0){40}}
\put(40,10){\line(-1,-1){20}}
\thicklines
\put(20,-25){\line(0,1){30}}
\put(60,-25){\line(0,1){30}}
\put(40,-5){\line(0,1){30}}
\put(80,-5){\line(0,1){30}}
\put(20,-20){\vector(0,1){7}}
\put(60,-20){\vector(0,1){7}}
\put(40,0){\vector(0,1){7}}
\put(80,0){\vector(0,1){7}}
\put(20,-10){\vector(0,1){11}}
\put(60,-10){\vector(0,1){11}}
\put(40,10){\vector(0,1){11}}
\put(80,10){\vector(0,1){11}}
\end{picture} \,.
\ee
The sums for $F^{01}C^2$ and $F^{20}C^1$ work similarly.

Alternatively, one can re-arrange this sum in terms of what $A$ fields
couple to each $C$ field;
\be
\sum_{x\in {\cal M}_3} F^{12} C^0 =
\sum_{x\in {\cal M}_3} \CL^0(x) \times \frac{1}{8}\sum_8 \FL^{12} =
\begin{picture}(60,30)
\thicklines
\put(40,-10){\line(0,1){30}}
\multiput(10,-20)(20,20){2}{\line(1,0){40}}
\multiput(10,10)(20,20){2}{\line(1,0){40}}
\multiput(10,-20)(40,0){2}{\line(1,1){20}}
\multiput(10,10)(40,0){2}{\line(1,1){20}}
\thinlines
\multiput(20,-10)(0,30){2}{\line(1,0){40}}
\multiput(30,-20)(0,30){2}{\line(1,1){20}}
\end{picture}
\ee
Note that $F^{12}$ involves a signed sum, see \Eq{FL_is}, so the
``middle'' lines in the field strength squares above cancel off, the
contribution associated with one $C^0$ field ``link'' is,
\be
F^{12} C^0 \rightarrow
\begin{picture}(60,30)
\thicklines
\put(40,-10){\line(0,1){30}}
\multiput(10,-20)(20,20){2}{\line(1,0){40}}
\multiput(10,10)(20,20){2}{\line(1,0){40}}
\multiput(10,-20)(40,0){2}{\line(1,1){20}}
\multiput(10,10)(40,0){2}{\line(1,1){20}}
\multiput(10,-20)(0,30){2}{\vector(1,0){25}}
\multiput(70,0)(0,30){2}{\vector(-1,0){25}}
\multiput(50,-20)(0,30){2}{\vector(1,1){12}}
\multiput(30,0)(0,30){2}{\vector(-1,-1){12}}
\end{picture}
\ee

\bigskip

In the continuum, time evolution requires fixing initial field values
$\phi,A,C$ and their time derivatives $D_0 \phi, F_{0i}, H_{0i}$.  On
the spatiotemporal lattice, we fix the links on two initial time slices,
$t=0$ and $t=a_t$.  The choice is constrained by Gauss' law, which is
the extremization of the lattice action with respect to a temporal link
$\AL^0(x)$ or $\CL^0(x)$.  Extremization of the action with respect to the
$\phi$ and $A$ variables on the $t=a_t$ layer fixes the values on the
$t=2a_t$ layer once we choose values for the temporal links (choose the
time dependent gauge), which is most conveniently done by setting
$\AL^0=\CL^0=0$.  In the $c_{\rm cs}=0$ theory, there is a 1-1
correspondence between equations of motion and $t=2a_t$ variables;
variation of $\AL^i(x,a_t)$ determines $\AL^i(x,2a_t)$ explicitly.  At
finite Chern-Simons term, the relations between the $t=a_t$ equations of
motion and the $t=2a_t$ fields are implicit and must be solved
iteratively by perturbing in $c_{\rm cs}$.%
\footnote{%
    Specifically, the Chern-Simons term means, for instance, that the
    variation of the action with respect to $\AL^i(x,a_t)$ depends on
    $\AL^j(x,2a_t)$, which is to be determined.  One starts by guessing
    that $\AL^j(x,2a_t)=\AL^j(x,at)$ to evaluate the equation of motion
    at $x,a_t$.  This gives a first guess for $\AL^i(x,2a_t)$ (and
    simultaneously, all other links at time $2a_t$).  This guess is used
    to compute the Chern-Simons term, redetermining the equation of motion
    at time $a_t$ and a better guess for the variables at $2a_t$.  The
    process is iterated until numerical convergence, typically of order
    4-7 iterations.
    }
This works because the
Chern-Simons term has one derivative, while the $F_{0i}^2$ term has two;
so the iteration converges in powers of $a_t/a$.

We should also verify that the action used is gauge invariant.
Changing the gauge at point $x$ means shifting $\phi(x) \rightarrow
e^{-i\theta} \phi(x)$ and making a change in the gauge fields which will
keep the $\phi$ field derivative term unchanged.  Examining
\Eq{phi_latt}, we see that this requires, 
\baray
\phi(x)  & \rightarrow & e^{-i\theta} \phi(x) \\
\AL^i(x) & \rightarrow & \AL^i(x) + \theta \\
\AL^i(x{-}a\hat{i}) & \rightarrow & \AL^i(x{-}a\hat{i}) - \theta
\earay
so each link starting at $x$ is shifted by $\theta$ and each link ending
at $x$ is shifted by $-\theta$.  This clearly cancels in the field
strength, $F^{12}$; each corner of the square in \Eq{FL_is} has one link
entering and one link leaving.  This ensures the gauge invariance of the
$F_{ij}^2$ term in the action, and the gauge invariance to changes in
the $A$ field gauge of the $F\wedge C$ term.  What about $C$ field gauge
changes in the Chern-Simons term?

\begin{figure}[thb]
\centerbox{0.3}{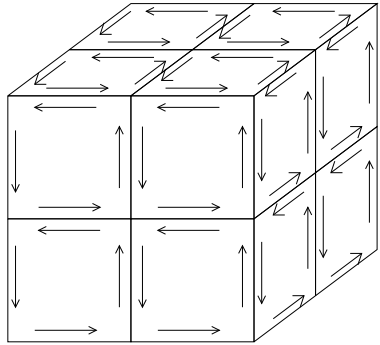}
\caption{\label{fig_box}Magnetic fields contributing to the variation of
  the action with respect to a gauge change in the $C$ field at the
  center of the cube.  This sum
  of $A$ field links exactly cancels--a lattice implementation of a
  ``boundary of a boundary.''}
\end{figure}

To see that the Chern-Simons term is invariant to $C$ field gauge
change, first note what such a gauge change adds to the action.  Each
$C$ link entering or leaving the site $x$ is shifted by $\pm \theta$
($+$ if it leaves the site).  This contributes $\theta$ times a signed
sum of $\FL^{\mu\nu}$ ``squares'' to the action.  But when we look at
the signed sum in detail, see Figure \ref{fig_box}, we find that it is
the surface of a box, such that each link appears in two $\FL^{\mu\nu}$,
with opposite orientation; the sum cancels.

\end{document}